# Rebuttal to "Comment by V.M. Krasnov on 'Counterintuitive consequence of heating in strongly-driven intrinsic junctions of $Bi_2Sr_2CaCu_2O_{8+d}$ Mesas' "

C. Kurter,[1,2] L. Ozyuzer,[1,3] T. Proslier,[1,2] J. F. Zasadzinski,[2] D. G. Hinks,[1] and K. E. Gray[1],[*]

[1]*Materials Science Division, Argonne National Laboratory, Argonne, Illinois 60439, USA*
[2]*Physics Division, BCPS Department, Illinois Institute of Technology, Chicago, Illinois 60616, USA*
[3]*Department of Physics, Izmir Institute of Technology, TR-35430 Izmir, Turkey*

In our article [1], we found that with increasing dissipation there is a clear, systematic shift and sharpening of the conductance peak along with the disappearance of the higher-bias dip/hump features (DHF), for a stack of intrinsic Josephson junctions (IJJs) of intercalated $Bi_2Sr_2CaCu_2O_{8+\delta}$ (Bi2212). Our work agrees with Zhu et al [2] on unintercalated, pristine Bi2212, as both studies show the same systematic changes with dissipation. The broader peaks found with reduced dissipation [1,2] are consistent with broad peaks in the density-of-states (DOS) found among scanning tunneling spectroscopy [3] (STS), mechanical contact tunneling [4] (MCT) and inferred from angle (momentum) resolved photoemission spectroscopy [5] (ARPES); results that could not be ignored. Thus, sharp peaks are extrinsic and cannot correspond to the superconducting DOS. We suggested that the commonality of the sharp peaks in our conductance data, which is demonstrably shown to be heating-dominated, and the peaks of previous intrinsic tunneling spectroscopy (ITS) data implies that these ITS reports might need reinterpretation.

Parts of the Comment by Krasnov (6) address our paper and other parts defend the absence of heating in his previous works [7,8]. The latter is relevant since our paper implies that the previously identified sharp conductance peaks for ITS would be a signature of heating. We begin with issues for our paper that include: (a) our use of the 'correct' spectra from MCT; (b) momentum conservation; and (c) our claim that the DHF is only found in ITS data with significantly reduced heating. Our latter statement (c) is a matter of incontrovertible published data [2,9,10], although the Comment author incorrectly adds four and deletes one of these references in *his* list. The issue of momentum-dependence of tunneling (b) is a relevant issue that is not fully resolved. However, any claim that the sharp peaks result from coherent tunneling [11], uniquely in ITS structures, has been experimentally ruled out by our shortest mesa (N=6 IJJs) data. Those data exhibit a significantly broadened peak, while the alignment of axes in those junctions cannot be different than the taller mesas with sharp peaks (also see the same behavior in Ref. 2 for pristine Bi2212). That crucial, but *straightforward, conclusion* is clearly addressed in our paper, while the Comment author claims that we have "discarded (it) in passing". Furthermore, momentum-dependent ARPES data [5] show a broadened peak at the anti node and flat spectra at the node, but never a sharp peak. We justify our use of the 'correct' spectra from MCT (a) by the universal agreement of MCT, STS and APRES for a broadened DOS in Bi2212, while the author presents no such confirmation of the sharp peaks in ITS. As such, we stand by the conclusions stated in our article [1].

Interestingly, Krasnov's synopsis of our work [6] states: "Nevertheless, they confirmed that self-heating is reduced in smaller mesas and even observed some reasonable spectroscopic features for their smallest mesa, although at a reduced voltage." This statement totally negates the premise of the Comment since what Krasnov refers to as "reasonable" corresponds to relatively broader conductance peaks and a clear DHF. Thus the ITS data emulates our apparently *unreasonable* sharp peaks, which resulted from increased self-heating in our largest mesas!

The Comment author is emboldened to criticize our paper by consistently ignoring the reduction of dissipation per unit volume by a factor of 20 through our use of intercalated Bi2212. Thus he fixates on the fact that our mesas have larger areas (100 µm$^2$) compared to the ITS literature (mostly >10 µm$^2$), even though the ten-times larger area is more than compensated by the twenty-times lower specific dissipation, i.e., our total power is roughly half as large and it is distributed over a larger area. Also, the very small ITS mesas (0.16 to 12 µm$^2$) of Zhu, et al [2] are in excellent agreement with our interpretation.

Although not intended to be, the parts of the Comment [6] defending the lack of heating in previous ITS data are the basis for reinterpreting some of these papers.

Previous works of the Comment author [7,8] argue that the temperature rise, ΔT, is negligible and thus they contend that their sharp peaks are intrinsic to Bi2212. As the two claims, (a) that the sharp peaks are intrinsic [6-8] due to negligible heating in ITS and (b) that the peaks progressively sharpen as the effect of heating increases [1,2], are incompatible, we address the evidence for negligible ΔT in ITS mesas [6-8] in the next section. Based on this discussion, we stand by our conclusion that it is the I(V) of the heated mesa that is the best measure of its temperature and the ITS data look convincingly similar to both our taller mesa data, that demonstrably result from significant heating, and the data of Ref. 2 for their largest-area mesa in which greater heating is expected.

**Heating tests in previous ITS publications**

Upon re-examination, all the arguments [6] for negligible ΔT are weak, at best (see below), except for one. The strongest evidence for the Comment author's claim of negligible ΔT is the uniformity of the sharply peaked dI/dV data (peak position and shape) in Fig. 2a [8] for differing mesa areas (2-25-μm$^2$). Based on theoretical models [8,12], for heat removal *through the substrate*, this area independence would only occur by a heating scenario if the linear mesa dimensions are all >200 μm, and so he reasonably concludes a negligible ΔT. However, the uniformity of dI/dV versus mesa area is in direct conflict with Ref. 2 that shows an areal dependent I(V) with backbending seen for a mesa of area 12-μm$^2$ in its Fig. 5a. Also, the conclusion that sharp peaks are intrinsic seems impossible to reconcile with the data of others showing (a) a transition from similar appearing sharp peaks in dI/dV to broader peaks *as the dissipation is reduced* [1,2] or (b) the broader peaks that appear universally in non-driven experiments *with negligible dissipation and heating*, e.g., in ARPES [5], MCT [1,4] and STS [3] data. Perhaps these theoretical heating models [8,12] are incomplete, e.g., by over simplifying the boundary condition for the underlying crystal at the base of the mesa or by ignoring thermal gradients in the mesa. In view of the collected evidence, it would seem incumbent on the Comment author to explain (a) the transitions from sharp to broader peaks with reduced dissipation [1,2]; (b) the broad peaks seen universally with negligible dissipation

[1,3-5]; and the backbending seen in nominally identical mesa structures of area 12-μm$^2$ [2].

To go beyond the Comment author's calculation, a complex ITS structure was fabricated in which the resistance of a close neighboring mesa [8] was used to estimate the dissipating mesa's temperature (see Comment item iii).  In a highly driven state the attainment of thermal equilibrium between these mesas is perhaps complicated.  Importantly, one mesa is driven with a high current, while the other is not.  Both mesas have gold films on top to provide some cooling.  Why should they be at the same temperature?  A valid thermometer is only thermally anchored to the sample with no other source of heat (or cooling).  But by the author's analysis the 'thermometer' mesa would be cooled by its Au film and would certainly be at a lower temperature.  Even if the author's analysis was quantitatively wrong (see section on heat removal below), the Au film will provide some cooling and there is negligible dissipation in the thermometer mesa.  It is also important to realize that the thermometer mesa's resistance can only be correlated to an *average* over its temperature profile.  In view of its potential thermal gradient, from the top Au-film heat sink and the negligible internal dissipation, the colder (highest resistivity) parts of the thermometer mesa, i.e., nearest the Au film, will dominate its measured resistance.  In that regard, also note that in their data (Fig. 3a of Ref. 8), the temperature is not a single valued function of resistance, and to compound this there is very little difference, in the equilibrium thermometer resistivity between 25 and 75 K.  So, their conclusion [8] that the temperature at the base of the thermometer mesa is only 24.8 K, for the quoted heater current of 0.4 mA, may be greatly in error.  Surely the *average* (quoted as 24.8 K, but equally consistent with 75 K) will be less than the hottest part at the base of the thermometer mesa.  Thus, both the base temperature and the derived heat transfer coefficient (Fig. 3b of Ref. 8) must be expected to be larger, possible by factors of 2-3.  Hence, any conclusion based on the 'thermometer' resistance has significant uncertainties.

An earlier attempt to argue against heating in the ITS data is in the Comment author's paper [7] showing the constancy of the quasiparticle branch spacing.  To test its validity, one should compare the maximum power dissipated in the highest quasiparticle branch of Fig. 3a in Ref. 7 (~90 µW) with that of the same I(V) at the conductance peak (~1000 µW), and then the claim seems suspicious. The latter value is ten times larger and the point of return to the 'normal state' conductance is yet larger, being ~1800 µW. If we assume that the mesa temperature reaches $T_c$~90 K upon return to the 'normal state', then ΔT for the branch with the highest power can be estimated to be ~5 K to bring the mesa temperature to ~10 K. Such an increase is not expected to have a measureable effect on the energy gap or the branch spacing, in spite of the mesa temperature being 90 K in the 'normal state'. Thus, this is hardly a convincing argument.

Finally, item (viii) of the Comment attempts to imply that the cusp-like conductance peak cannot represent the transition of the mesa to the normal state (with $T>T_c$) by pointing out the phonon resonances occur up to the sharp peak [13]. There is, of course, no inconsistency with our interpretation since $T<T_c$ everywhere in the mesa below the peak and the mesa is fully normal only after the conductance drops to the normal state value at even higher voltages than the peak.

In summary, most of the above points imply that the basis for Comment author's argument against heating in ITS data is very weak, at best, while the direct conflict with Ref. 2 for the remaining one, the independence of the I(V) on mesa area, has not been addressed. We feel that it is the I(V) of the heated mesa that is the best measure of its temperature and the ITS data look convincingly similar to our data that is demonstrably a result of significant heating.

**Heat removal in mesa configurations**

The Comment author's conclusion of negligible heating starts with their calculation for the spread of heat via the Au film atop the ITS mesa that implies it dominates heat removal [12]. This Au film continues beyond the mesa area, being supported by a $CaF_2$ insulating layer, and their previous analysis [12] claims the Au film-$CaF_2$ structure "can be an order of magnitude less" thermal resistance than the path directly through the base of the mesa into the underlying Bi2212 crystal. Unfortunately, the calculation of this auxiliary pathway neglected the Kapitza-like thermal boundary resistances, $R_K$, due to the acoustic impedance mismatches [14] between (a) the Bi2212 mesa and Au top film; (b) the Au film and the $CaF_2$ layer; and (c) between the $CaF_2$ layer and Bi2212 crystal.  These could play dominant roles for such thin, 100-nm $CaF_2$ and Au layers.  It is instructive to consider the known $R_K$ data [15] for In-$Al_2O_3$ interfaces that should be a reasonable approximation of the acoustic mismatch at the Au/$CaF_2$ interface.  The extrapolated $R_K$ is two orders-of-magnitude higher than the thermal resistance calculated for the $CaF_2$ layer [12] at 25 K and even higher at lower T.  The adversity of this comparison, plus the need to add two additional boundary resistances (albeit likely to have smaller acoustic mismatches), implies the existing calculation alone cannot prove that this cooling path is dominant.  Another thorny issue is the difficulty of reconciling this conclusion with the sub-µm mesa results [2] that used a virtually identical structure with the 50-100-nm gold film on $CaF_2$ for potential cooling.  These data [2] would imply that the Au film is ineffective when the mesa area exceeds 1-µm², and the Comment author needs to address this issue.  Otherwise one might conclude from the above that the dominant heat-flow path may commonly be directly from the mesa to the underlying Bi2212 crystal.

In our case [1] a bulk, 100-micron diameter, sharpened Au wire forms a metal-to-metal contact to the Au film atop the mesa at some random point.  This shunts the heat directly to a heat-sunk, bulk Au wire.  There are no additional Kapitza-like thermal boundary resistances beyond that between the Bi2212 mesa and Au top film that is common to all mesas.  In our paper, the empirical thermal resistance is ~70 K/mW.  Inexplicably, the Comment author states this "provides a much worse heat sink" whereas their claimed values at 4.2 K [8] are 25-60 K/mW, and we have shown above that their claims [8] are quite possibly significant underestimates.

**Dissipation vs. mesa area**

The Comment author agrees with us that heating effects can be minimized by the use of smaller mesas, more effective heat removal, or short pulses and those ideas are clearly indicated in our paper, but he consistently ignores the reduction of specific dissipation through our use of intercalated Bi2212.  The Comment author concludes that another paper [9] with an intercalated mesa "does not have 'greatly minimized heating effect'", because the area was 200 µm², again ignoring the effect of intercalation. To continue, a better understanding of the effects of mesa area is also needed.

The Comment author suggests that *compared to our mesas*, mesa areas and dissipated powers in ITS have been reduced more than four hundred times in recent years. The implication is that we have heating and the others do not. To highlight smaller areas, he notably references sub-µm area mesas of Ref. 2 and those made by Latyshev [15] (who used a focused-ion-beam on Bi2212 whiskers with no Au film for added cooling), plus his own ~3x3-µm$^2$ mesas. Although some mesa areas [2,15] are 100-400 times smaller than ours, Ref. 2 shows that the beneficial effects of reduced area are only found when the linear dimension of the mesa drops below a thermal healing length of ~1 µm. The Latyshev data [15] for a mesa of area 1.5-µm$^2$ show scant evidence for the ubiquitous dip feature in the Bi2212 spectrum and a narrow peak width parameter (see [1]) of ~0.15, thereby indicating significant heating. This data is reasonably consistent with our N=12, IJJ stack which is dominated by heating. Thus the smaller area does not overcome the factor of ~40 larger power per unit area [15] due to the use of unintercalated Bi2212 and taller mesas (we assumed an effective height, N, of 25, not the actual N=50, since cooling occurs equally from both sides). It seems unlikely that even the use of ~0.5 micron linear dimension could completely compensate these factors. The Comment author's other references to this point are his own papers that use mesas down to ~10-µm$^2$ and stack heights of N~10. As pointed out above, these dissipate about twice the total power of our smallest mesas (due to our twenty-times lower specific dissipation). Thus we cannot comprehend the Comment author's portrayal of a great reduction of heating in recent research obtained by "400 times smaller areas and powers" compared to our mesas.

In summary, the principal shortcomings of reasoning include (a) ignoring the reduction of specific dissipation in intercalated Bi2212 and (b) fixating on the effects of mesa area on heating. The problematic reasoning for (b) starts with an uncertain calculation (above) of heat removal by the top Au film and ignores both the 1-µm thermal healing length [2] and the reduction of specific dissipation by intercalation [1,9].

**Summary**

We feel that this Comment [6] lacks credibility and authenticity. For example, the Comment author points out the large Josephson current ($I_cR_N$ product) in MCT data [18], and claims this proves coherent tunneling. The logic here is confusing as Yamada and Suzuki [11] directly connect the sharp conductance peak to coherent tunneling and, instead, Ref. 18 exhibits a broadened peak. Further, we explain why the criticisms of our work in the Comment lack merit and we show that the arguments in the Comment author's papers against heating display significant uncertainty. We have addressed all of the relevant issues raised in the Comment. What the Comment author would need to do is to explain (a) the transitions from sharp to broader peaks with reduced dissipation [1,2]; (b) the broad peaks and strong DHF seen universally with negligible dissipation [1,3-5]; and (c) the backbending seen in mesa structures [2] nominally identical to their own. The Comment author states, "any universal shape for all tunneling characteristics, irrespective of experimental details, should not be expected" but in fact a high degree of reproducibility and consistency has now been demonstrated with STS, MCT and a few ITS experiments where heating was truly minimized. That the DOS from these experiments are consistent with momentum-resolved ARPES cannot be ignored. Thus the shape of the ITS conductance is the best indicator of self-heating (not heating models or uncertain thermometry) and we reiterate our claim that sharp conductance peaks and absent (or weak) DHF do not correspond to the equilibrium superconducting DOS.

Work supported by UChicago Argonne, LLC, operator of Argonne National Laboratory, a U.S. Department of Energy, Office of Science Laboratory, operated under Contract No. DE-AC02-06CH11357 and TUBITAK (Scientific and Technical Research Council of Turkey) Project No. 106T053. L.O. acknowledges support from Turkish Academy of Sciences, in the framework of the Young Scientist Award Program (LO/TUBA-GEBIP/2002-1-17).